\documentclass[preprint,showpacs,preprintnumbers,amsmath,amssymb,superscriptaddress]{revtex4-1}
\usepackage{graphicx}
\usepackage{dcolumn}
\usepackage{bm}

\begin{document}

\title{Cross sections for electron collision with pyridine [C$_5$H$_5$N] molecule.} 
 
\author{Czes{\l}aw Szmytkowski}%
\affiliation{
Department of Atomic, Molecular, and Optical Physics, 
Faculty of Applied Physics and Mathematics, 
Gda\'{n}sk University of Technology,  
ul.~G.~Narutowicza~11/12, 
80-233~Gda\'{n}sk, Poland
}
\author{Sylwia Stefanowska}%
\affiliation{
Department of Atomic, Molecular, and Optical Physics, 
Faculty of Applied Physics and Mathematics, 
Gda\'{n}sk University of Technology, 
ul.~G.~Narutowicza~11/12, 
80-233~Gda\'{n}sk, Poland
}
\author{Natalia Ta{\'n}ska}
\affiliation{
Department of Atomic, Molecular, and Optical Physics,  
Faculty of Applied Physics and Mathematics, 
Gda\'{n}sk University of Technology, 
ul.~G.~Narutowicza~11/12, 
80-233~Gda\'{n}sk, Poland
}
\author{Bo{\.z}ena {\.Z}ywicka}
\affiliation{Nicolaus Copernicus High School No.~I, ul.~Wa{\l}y~Piastowskie~6, 80-855~Gda{\'n}sk, Poland
}
\author{El{\.z}bieta Ptasi{\'n}ska-Denga}%
\affiliation{
Department of Atomic, Molecular, and Optical Physics,
Faculty of Applied Physics and Mathematics, 
Gda\'{n}sk University of Technology, 
ul.~G.~Narutowicza~11/12, 
80-233~Gda\'{n}sk, Poland
}
\author{Pawe{\l} Mo{\.z}ejko}%
\email{paw@pg.edu.pl}
\affiliation{
Department of Atomic, Molecular, and Optical Physics, 
Faculty of Applied Physics and Mathematics, 
Gda\'{n}sk University of Technology,  
ul.~G.~Narutowicza~11/12, 
80-233~Gda\'{n}sk, Poland
}

\begin{abstract}
The absolute \textit{grand}\,-total cross section (TCS) for electron scattering from pyridine, C$_5$H$_5$N, molecules has been measured at impact energies from 0.6 to 300~eV in the linear electron-transmission experiment. 
The obtained TCS energy dependence appears to be typical for targets of high electric dipole moment; 
the cross section generally decreases with rising energy, except for the 3--20~eV range, where a broad enhancement peaked near 8.5~eV is clearly visible. 
Below 10~eV, some weak structures which can be attributed to resonant scattering processes are also discernible. 
The present TCS energy dependence is compared with TCS experimental data obtained recently for energies above 13~eV by Traor{\'e} Dubuis et al. \cite{Trao18}. 
The TCS for pyridine has been confronted with TCS for benzene to search how the replacement of the CH group in the benzene ring with the nitrogen atom influences the electron scattering process.   
In addition, for pyridine and its halogenated derivatives: 2-chloropyridine [2-C$_5$H$_4$ClN] and 2-bromopyridine [2-C$_5$H$_4$BrN], integral elastic (ECS) and ionization (ICS) cross sections have been calculated at intermediate and high electron-impact energies within semiempirical approaches. 
For pyridine the sum of ECS and ICS is in reasonable agreement with the measured TCS above 40~eV. \\

{\bf keywords}:
electron scattering; pyridine; total cross section; resonant electron scattering
\end{abstract}

\maketitle

\section{Introduction}
Electron scattering data are strongly desirable and very useful in many scientific areas and technological applications \cite{Sanc09,Maso14,Altw16,Chri04}. 
In the last decades increasing attention has been devoted to understanding elementary phenomena that accompany the interaction of electrons with biologically relevant molecules \cite{Brun17,Kohanoff17}.  

Pyridine [C$_5$H$_5$N] is an aromatic heterocyclic compound structurally related to benzene [C$_6$H$_6$], with one CH group in the ring replaced by a nitrogen atom (cf. Fig.~1). 
The pyridine unit occurs in many compounds of biological importance, for example, in nicotine and B-group vitamins. 
Also, pyridine is used as a precursor to agrochemicals and pharmaceuticals and as a reagent and solvent. 
Numerous pyridine derivatives have importance for modern clinical applications \cite{Alta15}. 

Study on the electron scattering from the pyridine molecule has  quite a long tradition. 
Early experiments focused on the study of temporary negative ion states formed at low impact energies \cite{Hueb68,Pisa73,Nenn75,Math76,Mode83}; 
the electron induced electronic transitions in the gas-phase pyridine \cite{Jons69,Doer72,Walk89}; 
and the electron-impact ionization efficiency near the threshold \cite{Arim84}. 
Electron transmission experiment with thin film of solid target  suggested the formation of compound states also in solid pyridine \cite{Sanc79}. 
The features observed in the aforementioned works were located on the energy scale, but the intensities of the investigated processes were given in arbitrary units only what makes such results inconvenient for practical applications. 
Renewed interest in the electron scattering with biomolecules, among them in pyridine, appeared after observation that low-energy electrons can lead to the break-up of DNA \cite{Boud00}. 
Formation of resonant states in the electron-pyridine scattering was examined also in computations using the Schwinger multichannel method \cite{Barb13} and R-matrix method \cite{Sier14}. 
Cross sections for electron induced ionization of pyridine were calculated \cite{Bull14,Gupt14,Sier14} and measured in absolute scale \cite{Jiao06,Bull14}. 
Examination of the measured electron energy-loss spectra allowed to observe and assign the triplet excited states of pyridine \cite{Line16}. 
Recently, the formation of anionic species resulting from the dissociative attachment of low-energy electrons to pyridine has been investigated \cite{Rysz17}. 
Just in the course of the present experiment, the absolute TCS data have been published \cite{Trao18}; measured in the transmission experiment from 13 to 902~eV and computed between 1 and 1000~eV. 

\begin{figure}[h]
\begin{center}
\includegraphics[width=16cm,height=12cm,angle=0]{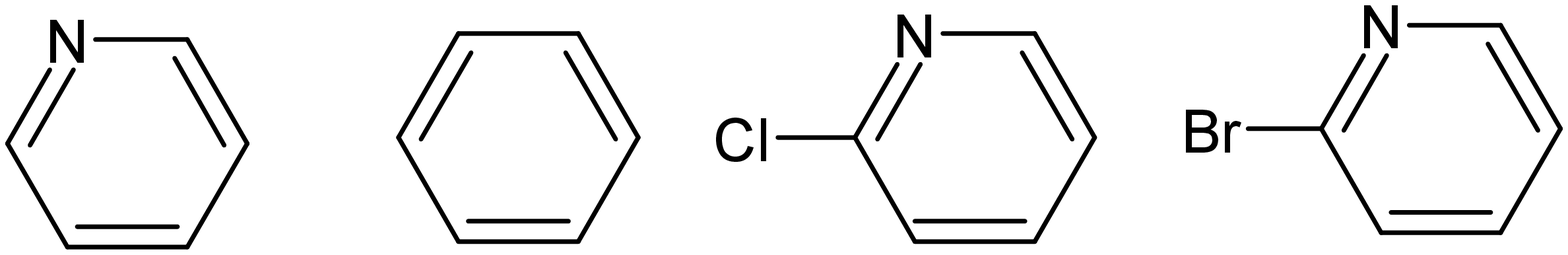}
\caption
{Structure of molecules studied. From left to right: pyridine, benzene, 2-chloropyridine and 2-bromopyridine.  
}
\end{center}
\end{figure}

Deficiency of absolute electron-scattering cross sections for the pyridine molecule prompted us to measure reliable absolute $grand$-total cross section (TCS) for this compound, for impact-energy range from low to intermediates. 
Total cross section is that quantity describing the electron scattering which can be obtained without any normalization procedure, with a good accuracy over a wide energy range. 
Therefore, it may be used as a calibration standard or the upper limit for the normalization of particular scattering quantities, taken only in arbitrary units, as well as for the estimation of that quantities which are difficult to obtain. 
The TCS may serve also as one of the ranges of experimental tests of the reliability of theoretical models and computational procedures used in the electron-scattering calculations.

Present absolute \textit{grand}\,-total cross sections (TCS) were measured at electron-impact energies ranging from 0.6 to 300~eV using the linear electron-transmission method. 
To our knowledge, the experimental TCS data below 13~eV are not available in the literature. 
The observed low-energy features in our TCS were explained based on findings of previous experiments \cite{Hueb68,Pisa73,Nenn75,Math76,Mode83} and computations \cite{Barb13,Sier14}.   
The current electron-scattering TCS results for the pyridine [C$_5$H$_5$N] molecule were then compared to experimental TCS data \cite{Moze96,Gull98} for its isoelectronic 6-membered ring counterpart  benzene [C$_6$H$_6$]; the substituent effect is discussed. 
We present also elastic (ECS) and ionization (ICS) cross sections for pyridine and its two halogenated derivatives (2-chloropyridine and 2-bromopyridine) computed at intermediate and high electron-impact energies in the additivity rule approximation and the binary-encounter-Bethe approach, respectively.

\section{Experiment}
\subsection{\textit{Experimental procedure}}
The total cross sections for electron scattering from pyridine molecules reported in this work have been obtained using a linear electron-transmission method in single-collision conditions. 
The idea of the transmission method is based on the measurements of the attenuation of a projectile particle-beam passing through the scattering medium under study (for details see e.g. Ref.~\cite{Bede71}). 
The total cross section (TCS), $Q(E)$, for the scattering of projectiles of given energy $E$ from target particles, is related to the attenuation of the transmitted beam intensity through the Bouguer-de~Beer-Lambert (BBL) relationship: 
\[
I_{n}(E) = I_{0}(E)\; {\rm exp}[- Q(E)nL] .  
\]
\noindent 
Here, $I_{n}(E)$ is the measured intensity of the projectile beam after traversing a length $L$ of target medium whose the absolute number density is $n$, and $I_{0}(E)$ is the intensity of the beam taken in the absence of the target in the reaction volume. 

The experimental set-up and the measurement procedure used in the present electron-pyridine scattering experiment have been described in detail elsewhere \cite{Szmy97}, so only a brief outline is given here. 
A tunable-energy electron beam, formed in a system of electrostatic lenses coupled to an energy dispersing 127$^{\circ}$ electrostatic deflector, is directed into a reaction cell where its intensity is attenuated by the presence of the vapor sample under investigation. 
Those electrons which leave the cell through the exit aperture are energy discriminated by the retarding-field filter and eventually detected with the Faraday cup. 
Electron optics of the spectrometer is housed in a vacuum chamber evacuated to a base pressure of about 40~$\mu$Pa. 
The magnetic field along the whole electron trajectory is reduced below 0.1~$\mu$T with the system of Helmholtz coils. 

The quantities necessary for TCS derivation are taken directly in the present experiment and therefore cross section values reported in this work are given in absolute units, without any normalization procedure. 
$L$ was taken equal the distance (30.5~mm) between entrance and exit apertures of the reaction cell, 
while the target density value, $n$, is evaluated from the ideal gas formula corrected for the thermal transpiration effect \cite{Knud10} 
\[
n = \frac{p_{t}}{k \sqrt{T_{t}T_{m}}} ,  
\]
\noindent 
where: $p_{t}$ means the pressure of the vapor-target in the cell as measured by a capacitance manometer and 
$k$ denotes the Boltzmann constant; 
$T_{t}$ is the temperature of the target cell determined using a thermocouple;  
$T_{m}=322$~K~$ >T_{t}$ is the temperature at which the manometer head is held. 

The energy scale has been calibrated against the oscillatory structure visible around 2.3~eV in the transmitted current when molecular nitrogen was admixtured to the target under study. 
The declared inaccuracy of the energy scale ($\sim$ 0.1~eV) is higher than that resulting directly from the calibration due to the shift in energy, perceptible in the course of the long lasting experiment. 

A commercially supplied sample of  pyridine from Sigma-Aldrich, with a stated purity of 99.8\%, was distilled by freeze-pump-thaw repetitive cycles before use to remove volatile impurities. 
The target vapor was admitted into the spectrometer via a variable leak valve, alternately into the reaction cell and the outer vacuum volume, thus the pressure in the region of the electron optics was maintained constant (below 0.6~mPa) whether or not the target was present in the cell; 
that ensured a stable primary electron-beam intensity during both phases of the intensity measurements. 
Due to a low vapor pressure of pyridine at room temperature, the sample handling system has been maintained at elevated temperature about 315~K. 
The TCS measurements have been carried out at target-vapor pressures in the reaction cell between 70 and 200~mPa. 
Under these conditions no systematic variation of the measured TCSs with the target pressure was observed.  

\subsection{\textit{Uncertainty analysis}}
The final TCS value at each electron impact energy was derived as a weighted mean of results obtained in different runs. 
The statistical variations of the reported TCS for pyridine, estimated as one standard deviation of the weighted mean value from TCS values obtained in different runs, do not exceed 1\% below 100~eV and gradually increase up to nearly 2\%  at the highest electron-impact energies applied. 

However, the accuracy of the measured TCSs is mainly determined by the possible systematical uncertainties immanently associated with the transmission-type experiment \cite{Bede71}. 
The most serious problem arises due to inability to discriminate against electrons which are scattered elastically or with small energy losses through small angles in the forward direction and which contribute to the measured transmitted current, resulting in the lowering of the measured TCS \cite{Sull11}. 
The retarding field filter prevents only the electrons scattered inelastically in the forward direction to be detected together with those unscattered. 
The amount by which the experimental TCS might be lowered due to the \textit{forward-angle scattering effect} can be roughly estimated based on an angle distribution of the scattered electrons, measured or calculated. 
Taking into account theoretical elastic differential cross sections \cite{Barb13,Sier14,Trao18}, we found that the measured TCS can be lowered by 4--5\% around 3~eV, about 2--3\% within  30--100~eV, and 3--4\% above 200~eV. 

Another troublesome uncertainty in the electron-transmission experiment relates to unavoidable effusion of the target molecules through orifices of the reaction cell leading to: 
(i)  inhomogeneous target distribution, $n$, (especially in the vicinity of cell apertures) along the electron trajectory in the cell, and 
(ii) incorrect determination of the effective path length, $L$, of electrons across the sample volume. 
To estimate the uncertainty related to the factor $nL$ in the BBL formula we followed the method adopted from Ref.~\cite{Nels73}  to the present experimental conditions. 
The calculations show that both aforementioned effects nearly compensate and the estimated uncertainty of $nL$ is about 2--3\%, taking also into account the uncertainty in the pressure measurements. 
 
The overall systematical uncertainty in our absolute TCSs, estimated as the sum of potential systematic errors of all quantities taken in the experiment, amounts up to 9--11\% between 0.6 and 2~eV, decreasing gradually to 6--8\% within 2--5~eV, and to about 5\% between 5 and 100~eV, increasing again to 7--8\% at higher energies. 
It is to be noted that the reported TCS data are not corrected for the \textit{forward-angle scattering effect}. 
Finally, we note that the observed shift in energy may cause that the low-energy structures in the measured TCS become less pronounced, especially if they are located on steeply rising or falling side of the curve.

\section{Results and discussion}
\subsection{Pyridine, C$_5$H$_5$N}
Figure 2 shows the variation of absolute \textit{grand}\,-total electron-scattering cross section (TCS) for pyridine [C$_5$H$_5$N] measured in this work over the impact energy from 0.6 to 300~eV. 
A comparison is made with recent experimental TCS data obtained above 13~eV \cite{Trao18}.  
The numerical values of our TCSs are listed in Table I. 
 
\begin{figure}[h]
\begin{center}
\includegraphics[width=16cm,height=12cm,angle=0]{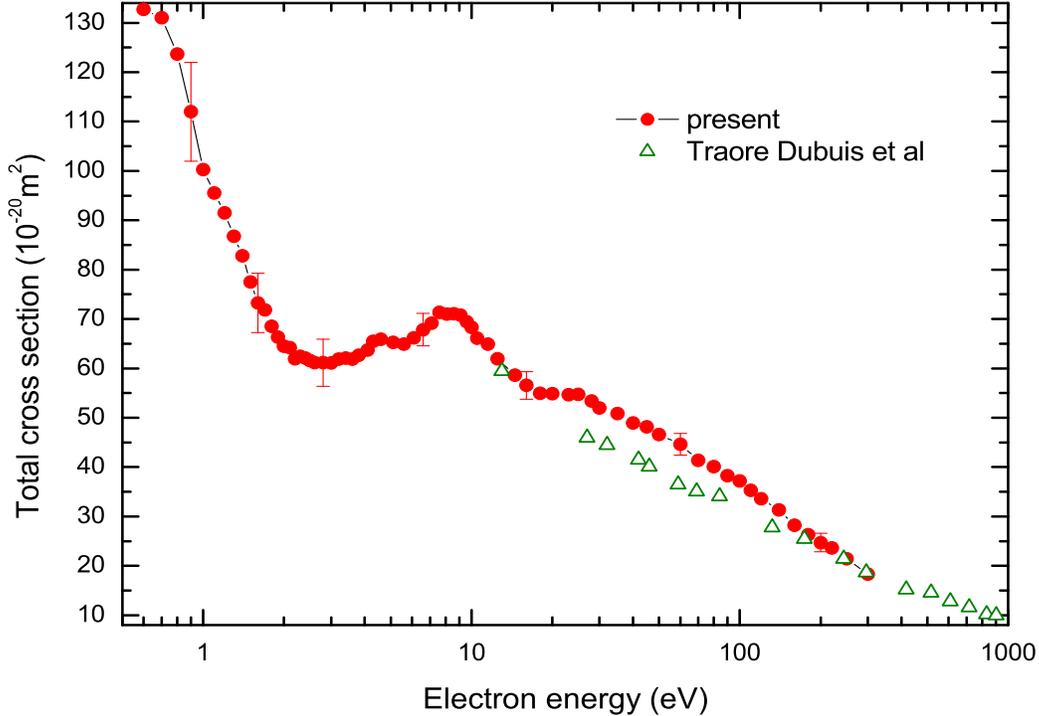}
\caption
{(Color online) Experimental total cross sections for the electron scattering from the pyridine (C$_5$H$_5$N) molecule:    
full (red) circles, present, error bars correspond to overall  experimental uncertainties;      
open (olive) triangles, from Ref.~\cite{Trao18}. 
}
\end{center}
\end{figure}

\begin{table}[h]
\caption{Absolute experimental electron-scattering total cross sections (TCSs) for the pyridine [C$_5$H$_5$N] molecule; in units of $10^{-20}$~m$^{2}$.}
\label{prop}
\begin{center}
\begin{ruledtabular}
\begin{tabular}{@{}*{10}{r}}

E (eV)&TCS  & E (eV)&	 TCS    &E (eV)&TCS  & E (eV)& TCS & E (eV)& TCS\\
\hline
0.6  &  133  &  2.0  & 64.5  &  4.3  & 65.5  & 11.5&  64.9 &   60 & 44.6\\
0.7  &  131  &  2.1  & 64.2  &  4.6  & 65.9  & 12.5&  61.9 &   70 & 41.4\\ 
0.8  &  124  &  2.2  & 61.9  &  5.0  & 65.3  & 14.5 &  58.6 &  80 & 40.1\\  
0.9  &  112  &  2.3  & 62.4  &  5.5  & 64.9  & 16 & 56.6 &  90 & 38.3\\    
1.0  &  100  &  2.4  & 62.1  &  6.0  & 66.2  & 18 & 55.0 & 100 & 37.2\\  
1.1  &  95.5  &  2.5  & 61.6  &  6.5  & 67.8  & 20 & 54.9 & 110 & 35.3\\ 
1.2  &  91.5  &  2.6  & 61.2  &  7.0  & 69.1  & 23  &  54.6 &  120 & 33.6\\
1.3  &  86.8  &  2.8  & 61.1  &  7.5  & 71.3  & 25  &  54.7 &  140 & 31.3\\ 
1.4  &  82.8  &  3.0  & 61.1  &  8.0  & 71.0  & 28  &  53.4 &  160 & 28.2\\ 
1.5  &  77.5  &  3.2  & 61.9  &  8.5  & 71.1  & 30  &  52.0 &  180 & 26.3\\
1.6  &  73.3  &  3.4  & 62.1  &  9.0  & 70.8  & 35  &  50.9 &  200 & 24.7\\
1.7  &  71.8  &  3.6  & 61.9  &  9.5  & 69.4  & 40  &  49.0 &  220 & 23.6\\
1.8  &  68.6  &  3.8  & 62.7  &  10   & 68.3  & 45  &  48.2 &  250 & 21.4\\
1.9  &  66.3  &  4.1  & 63.7  &  10.5 & 66.1  & 50  &  46.6 & 300  & 18.4\\
\end{tabular}
\end{ruledtabular}
\end{center}
\end{table}

The most distinct feature of the TCS energy curve for pyridine  (Fig.~2) is visible below 2~eV; 
at 0.6~eV, the lowest impact energy used, the magnitude of TCS reaches the highest value of above $130 \times 10^{-20}$~m$^{2}$ and then rapidly decreases to about $61 \times 10^{-20}$~m$^{2}$ in the vicinity of 2.8~eV, where TCS curve has its local minimum. 
The rapid rise in TCS towards low-impact energies may be in the predominant part due to strong permanent electric dipole moment of the C$_{5}$H$_{5}$N molecule ($\mu = 2.215$~D, see Table II). 
For polar targets the direct long-range  (point-charge---electric-dipole) forces are dominant and at low impact energies contributes significantly to the scattering -- the higher dipole moment reflects in more distinct increase of cross section towards thermal energies \cite{Itik77}. 
Because the C$_{5}$H$_{5}$N molecule has quite pronounced polarizability ($\alpha$ = $9.18 \times 10^{-30}$m$^3$, see Table II), some contribution to the TCS is also related to the interaction of electron with the induced electric dipole moment of target molecule.  
A closer inspection of the descending side of the TCS curve, below 2~eV, reveals that around 0.7 and 1.2~eV the TCS energy function changes slightly a slope. 
These two hardly perceptible TCS features are located in the energy range where distinct structures were visible in the experiments more suitable for the detection of weak variations of cross section \cite{Hueb68,Pisa73,Nenn75,Math76,Mode83}. 
These structures were attributed to the formation of two shape resonant states, $\pi^{*}_1$(b$_1$) and $\pi^{*}_2$(a$_2$), taking place when the incoming electron is temporarily accommodated on the lowest normally unfilled $\pi^{*}$ orbitals of the C$_5$H$_5$N molecule in its electronic ground state. 
Resonant maxima close to this energy range were also visible in the computed integral elastic cross sections \cite{Barb13,Sier14}. 
We suppose that two weak features visible in our TCS between 0.6 and 1.5~eV are the demonstration of these resonant processes.   

Above 3~eV the TCS energy curve shows a very broad enhancement peaking within 7.5 and 9~eV with the value of about $71 \times 10^{-20}$~m$^{2}$. 
On the low-energy side of this enhancement, between 4 and 5~eV, a weak hump in the TCS curve is perceptible. 
This feature corresponds to the 4--5~eV structure observed in the low-energy transmission experiments \cite{Nenn75,Math76,Mode83} and is related to the formation of the third resonant state, $\pi^{*}_3$(b$_1$), located around 4.5~eV. 
In the computed elastic cross sections \cite{Barb13,Sier14} that resonant structure appears at higher energy, within 5--6~eV. 
Worth noting is also some flattening of the TCS curve between 7 and 9.5~eV. 
In this energy regime, the transmission spectrum \cite{Math76} and anion yield curves \cite{Rysz17} also suggested the presence of some resonances. 

At electron impact energies above 10~eV, the TCS decreases systematically with energy increase down to about $18 \times 10^{-20}$~m$^{2}$ at 300~eV. 
Only around 25~eV some change in the slope of the TCS curve is discernible. 
The shoulder in this energy region of TCS is a feature which is common for complex hydrocarbons. 

Figure 2 shows also that, in the common energy range of compared  experiments, the present TCS results are in reasonable agreement with the recent data from Ref.~\cite{Trao18}, although some differences do exist. 
The most pronounced difference in the magnitude is visible between 30 and 60~eV where our TCS is higher by nearly 20\%, somewhat more than the combined declared uncertainties.

\subsection{Comparative studies}
\subsubsection{\textit{Experimental total cross sections for pyridine and benzene}}
In this section, we examine how the replacement of one CH group in the benzene ring with the nitrogen atom is reflected in the TCS energy dependence. 
For this purpose, in Fig.~3, the present experimental TCS for pyridine is compared with the experimental TCS curves for benzene: at very low energies the TCS was taken by Gulley et al \cite{Gull98}, and that above 0.6~eV was obtained in our laboratory \cite{Moze96}. 
The schematic geometry of both compared compounds is shown in Fig.~1; some their parameters are given in Table~II. 

\begin{figure}[h]
\begin{center}
\includegraphics[width=16cm,height=12cm,angle=0]{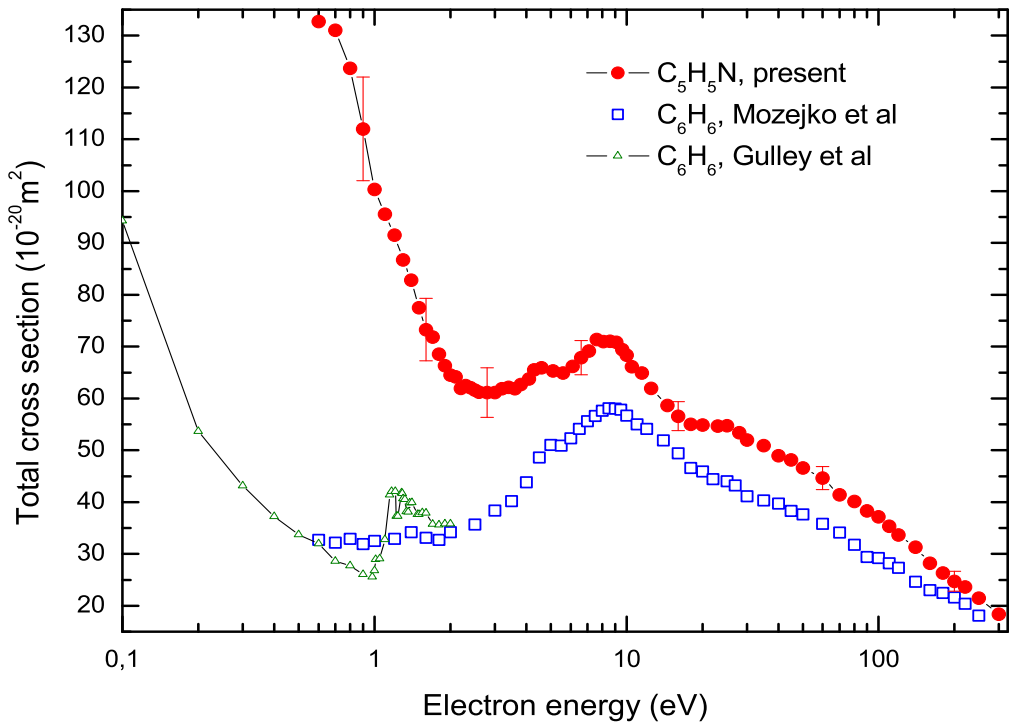}
\caption{(Color online) Comparison of experimental total cross sections for electron scattering from:   
pyridine (C$_5$H$_5$N), full (red) circles, present and  
benzene (C$_6$H$_6$), open (olive) triangles -- interpolation to guide the eyes, based on Ref.~\cite{Gull98}; open (blue) boxes, from Ref.~\cite{Moze96}. 
}
\end{center}
\end{figure}

Figure 3 shows, that according to the shape the TCS energy functions for benzene and pyridine look quite similar. 
Below 2~eV, the compared curves rapidly decrease with an energy increase, and above 3~eV, they have broad enhancement peaking near 9~eV. 
This similarity in the shape of TCS curves for benzene and pyridine is at low impact energies somewhat intriguing. 
For pyridine, one would expect a rapid rise in TCS toward the lowest energies because such behavior is rather typical for targets with high permanent electric dipole moment ($\mu_{pyridine} \sim 2.2$~D). 
On the other hand, for benzene (with $\mu \simeq 0$~D) early TCS  experiments \cite{Sueo88,Moze96,Mako03} indicated that slightly below 1~eV (up to 0.6~eV) the TCS is nearly constant, like for many nonpolar targets. 
However, the low-energy experiment \cite{Gull98} clearly showed that below 0.6~eV 
the TCS for C$_6$H$_6$ starts to increase rapidly with the energy decrease to the thermal energy region. 
Such low-energy TCS behavior in benzene might be attributed to the formation of parent negative ions with the lifetime of about 1~$\mu$s, what (like for the SF$_6$ molecule; cf. Ref.~\cite{Chri04}) would lead to large cross section at near zero energy.  
However, detailed study of electron attachment to benzene \cite{Fiel01} provides no evidence for the formation of relatively  long-lived benzene anions close to zero energy. 
Alternative explanation for such rapid rise in cross section for benzene is based on the virtual-state model scattering \cite{Gian98,Fiel01,Barb17}. 

Both TCS energy curves for benzene show one feature located between 1 and 2~eV related to electron capture into the lowest unfilled degenerate $\pi^{*}$(e$_{2u}$) orbital yielding temporary anion state in this energy regime (see Ref.~\cite{Barb17} and references therein). 
The replacement of the CH group by the nitrogen atom in the benzene ring removes this degeneracy and two shape resonant states are formed in pyridine, what reflects in the change of the slope of TCS curve near 0.7 and 1.2~eV. 

For benzene the TCS shows a shoulder between 4 and 6~eV, while a weak hump is located around 4.6~eV in the TCS for pyridine. 
Both these structures were also associated with the formation of resonant state, of mixed shape and core-excited character. 

With respect to the magnitude, the TCS for pyridine is generally higher than that for benzene. 
At 0.6~eV, the TCS for pyridine is larger by a factor of about 5 and that difference is mainly related to the direct interaction of the incoming electron with the polar pyridine molecule. 
For higher energies the difference in TCS magnitude systematically decreases and above 200~eV both TCS curves tend to merge.

\subsubsection{\textit{Calculated cross sections for pyridine [C$_5$H$_4$N] and its halogenated derivatives: 2-chloropyridine [2-C$_5$H$_4$ClN] and 2-bromopyridine [2-C$_5$H$_4$BrN]}}
Figure~4 shows integral elastic (ECS) and ionization (ICS) cross sections computed for pyridine and its derivatives in which one H atom (the next to N atom) was substituted with the Cl or Br atom: 2-chloropyridine [2-C$_5$H$_4$ClN] and 2-bromopyridine [2-C$_5$H$_4$BrN]; 
for the schematic geometry of those compounds see Fig.~1.

\begin{figure}[h]
\begin{center}
\includegraphics[width=16cm,height=12cm,angle=0]{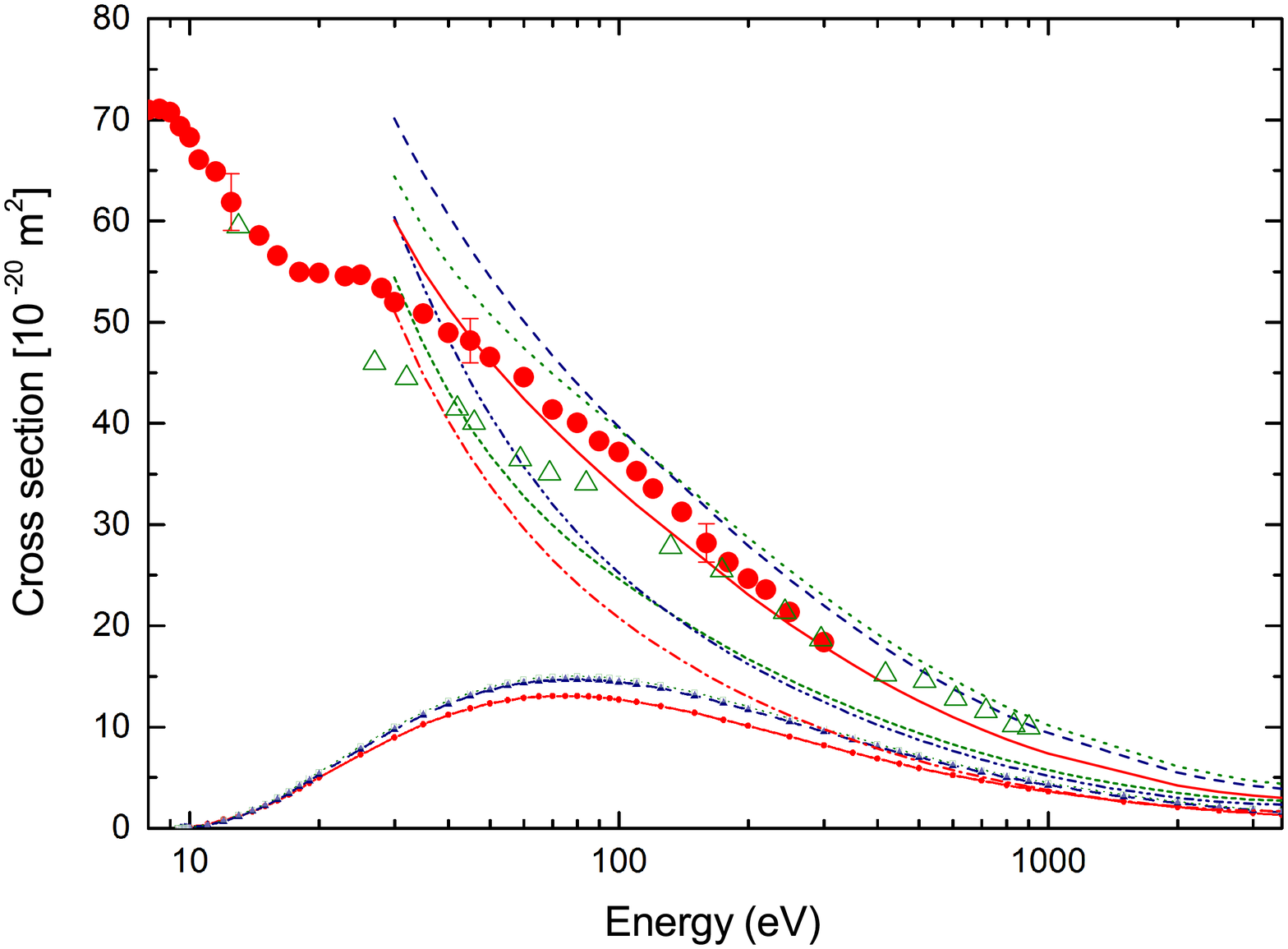}
\caption{(Color online) Cross sections calculated for pyridine and its halogenated derivatives. 
Elastic cross section: dash-dot (red) line, pyridine; dash-dot-dot (blue) line, 2-chloropyridine; short dash (green) line, 2-bromopyridine; 
ionization cross section: full (red) line with bullets, pyridine; dashed (blue) line with triangles, 2-chloropyridine; dotted (green) line with boxes, 2-bromopyridine; 
total: full (red) line pyridine; dashed (green) line, 2-chloropyridine; dotted (blue) line, 2-bromopyridine. 
Computed total cross section for pyridine is compared to respective  experimental data, full (red) circles, present, and open (green) triangles from Ref.~\cite{Trao18}. 
}
\end{center}
\end{figure}

The computed cross sections were obtained using simple methods: the ECS -- in the additive rule approximation \cite{Raj91} with the static+polarization interaction taken into account; while the ICS with the binary-encounter-Bethe approach \cite{Hwan96}. 
All quantities necessary in calculations have been obtained on HF and OVGF level with the Gaussian code \cite{Fris09} and 6-311++G(2d,2p) Gaussian basis set.  
The used theoretical approaches and computational procedures were described in detail in our previous works \cite{Moze12,Szmy18}, and therefore are not repeated here. 
For considered molecules the sum of ECS and ICS, which represents the computed total cross section, is also depicted in Fig.~4. 

Detail comparison of calculated ionization and total (\mbox{ECS+ICS}) cross sections for pyridine molecules with available theoretical and experimental data is shown in Figure~5. 
Our computed ionization cross section is in reasonable agreement with experimental data of Bull et al.~\cite{Bull14}.    
Although our data are lower than experimental results of Jiao et al.~\cite{Jiao06} and theoretical data~\cite{Gupt14,Trao18} they are still within declared combined experimental and computational uncertainties. 
For pyridine, above 40~eV, the computed total cross section (\mbox{ECS+ICS}) is in quite good agreement with our experimental TCS, it is also in reasonable accord with the high energy experimental and theoretical data from Ref.~\cite{Trao18}. 

\begin{figure}[h]
\begin{center}
\includegraphics[width=16cm,height=12cm,angle=0]{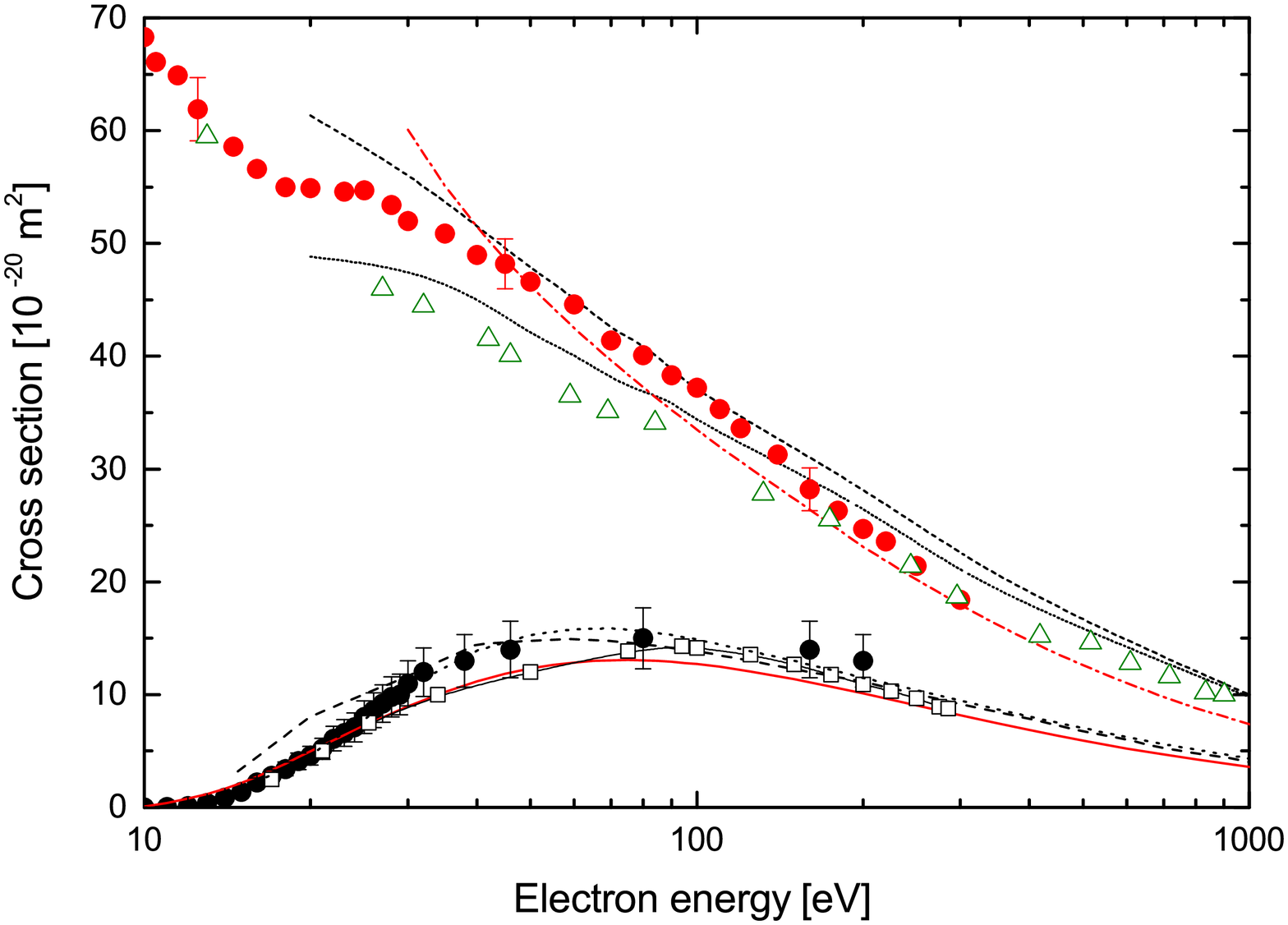}
\caption{(Color online) Comparison of cross sections calculated for pyridine molecule.  
Theoretical ionization cross section: full (red) line, present; dotted line~\cite{Gupt14};  dashed line~\cite{Trao18}. 
Experimental ionization cross section: full (black) circles~\cite{Jiao06}; full line with squares~\cite{Bull14}. 
Total cross section: full (red) circles, present experimental data; open (green) triangles experimental data from Ref.~\cite{Trao18}; dash-dot (red) line, present calculations; short dott line, IAM-SCAR calculations~\cite{Trao18}; short dash line, IAM-SCAR+I+R calculations~\cite{Trao18}. 
}
\end{center}
\end{figure}

\begin{table}[h]
\caption {Location of the low-energy resonant-like features,  $E_{\rm r}$, perceptible in the TCS curves compared in Fig.~3. 
Listed are also selected electric parameters for considered compounds (from Ref.~\cite{Lide95}): 
the permanent dipole moment, $\mu$, and the static dipole polarizability, $\alpha$. 
} 
\label{prop}
\begin{center}  
\begin{ruledtabular}
\begin{tabular}{@{}*{4}{l}}

                                   & $E_{\rm r}$          & $\mu$      &$\alpha$                \\ 
Molecule                    & (eV)                              &(Debye)    &($10^{-30}$m$^3$) \\
\hline
pyridine [C$_5$H$_5$N] &($\sim$0.7, $\sim$1.2, 4.6, 7--9.5)\footnotemark[1] & 2.215 & 9.18 \\
benzene [C$_6$H$_6$] & (1.15, 4.8, 8--9)\footnotemark[2] & 0 & 10.32 \\

\end{tabular}
\end{ruledtabular}
\footnotetext[1]{Present work.}
\footnotetext[2]{Reference~\cite{Moze96}.}
\end{center}
\end{table}

\section{Conclusions}
We have presented our experimental absolute total cross sections for electron scattering by the pyridine [C$_5$H$_5$N] molecule over wide energy range from 0.6 to 300~eV. 
TCS data measured at impact energies below 13~eV are presented for the first time. 
At low impact energies, the TCS rapidly decreases with energy increase and has a minimum located near 2.8~eV. 
Below the minimum, in the vicinity of 0.7 and 1.2~eV, a weak changes in the slope of the TCS energy curve are discernible. 
Above 3~eV the TCS energy dependence shows a very broad, highly asymmetric enhancement peaking between 7.5--9~eV.  
On the low-energy side of this enhancement, between 4 and 5~eV, a weak hump is clearly visible. 
Based on results of earlier low-energy experiments and calculations,  the TCS features observed around 0.7, 1.2 and 4.6~eV were attributed to the formation of short-living negative ion states.  
Our TCS energy dependence is in reasonable agreement with very recent TCS measurements \cite{Trao18}. 
To study how the replacement of the CH group in the benzene ring with the nitrogen atom reflects in the cross section energy dependence we have compared the TCS for pyridine and benzene  molecule. 

In addition, for pyridine and its halogenated derivatives: 2-chloropyridine [2-C$_5$H$_4$ClN] and 2-bromopyridine [2-C$_5$H$_4$BrN], integral elastic (ECS) and ionization (ICS) cross sections have been calculated at intermediate and high electron-impact energies in the additivity rule approximation and the binary-encounter-Bethe approach, respectively. 
For pyridine the sum of ECS and ICS is in good agreement with the measured TCS above 40~eV. 
That agreement suggests that our calculated \mbox{ECS+ICS} results can reasonably represent total cross section values also for 2-chloropyridine and 2-bromopyridine which may be useful for further applications. 
Monohalopyridines are considered as guest molecules in the preparation of superconducting crystals \cite{Prok15}.

\begin{acknowledgments}
This work has been supported in part by the Polish Ministry of Science and Higher Education (MNiSzW Project 2017-2018).  
Sylwia Stefanowska kindly acknowledge the support of the Polish Ministry of Science and Higher Education within the Diamond Grant program (Project no. DI2015 018945). 
Numerical computations have been performed at the Academic Computer Center (TASK) in Gda{\'n}sk.
\end{acknowledgments}

                     
\end{document}